\documentstyle[aps,multicol,prl,epsf] {revtex}
\draft
\begin{document}
\def \bm #1 {\mbox{\boldmath $#1$}}
\title { Estimation of initial conditions from a  scalar time series}
\author{ Anil Maybhate\footnote{e-mail: nil@prl.ernet.in}$^{1,2}$and
         R.E.Amritkar\footnote{e-mail: amritkar@prl.ernet.in}$^1$}
\address{$^1$Physical Research Laboratory,Navrangpura,Ahmedabad 380009 India}
\address{$^2$Department of Physics,University of Pune,Pune 411007 India}
\maketitle

\begin{abstract} 
We introduce a method to estimate the initial conditions of a mutivariable
dynamical system from a scalar signal. The method is based on a modified
multidimensional Newton-Raphson method which includes the time evolution
of the system. The method can estimate initial conditions of periodic and
chaotic systems and the required length of scalar signal is very small.
Also, the method works even when the conditional Lyapunov exponent is
positive. An important application of our method is that synchronization
of two chaotic signals using a scalar signal becomes trivial and
instantaneous.  
\end{abstract}

\begin{multicols}{2}

A trajectory that a given dynamical system traverses in its state space
depends on the particular set of initial conditions with which it starts.
In particular, the state of a chaotic system at a latter time is
exponentially sensitive to changes in its initial state~\cite{Dev}. This
defining feature of a chaotic system leads to a complex behaviour in state
space that appears random yet is deterministic, which means that an
initial state uniquely fixes the future course of its evolution. Though
their are several invariant measures of a chaotic system which are not
sensitive to the intial conditions, the exact trajectory crucially depends
on the intial state and hence is difficult to reproduce due to sensitivity
to initial conditions.

In light of these facts, it is interesting and important to ask whether
the complete set of initial conditions of a given multivariable dynamical
system can be estimated from a given scalar time series for a single state
space variable. We show that this question can be answered in the
affirmative and present a novel and simple method to estimate the initial
conditions. Our method is based on a modified multidimensional
Newton-Raphson method~\cite{Dev,PTVF} where we include the time evolution
of the system.  The length of the time series required for the
calculations is typically very small.

Our results raise some interesting issues regarding the information
content of a time series.  In standard embedding techniques, a vector
space is constructed from successive iterates of a single variable and a
trajectory is reconstructed in this space~\cite{ABST}. While embedding, it
is crucial to choose an appropriate time delay so that the successive
iterates are well resolved and contain qualitatively different
information. In our method we use a very small time series and the total
duration is typically much less than the standard delay time in embedding
techniques. It is interesting that we can recover the initial conditions
and hence the trajectory from such a small duration of time series.

An important application of our method is in the problem of
synchronization of two chaotic systems~\cite{PC1}. This problem itself has
attracted wide attention in recent times due to its potential application
to secured communication~\cite{CO,PKSP,JA} and parameter
estimation~\cite{Par,MA}. Our estimation of the initial conditions makes
the problem of synchronization almost trivial. We also find that our
method works for most of the cases where other methods fail~\cite{PC1,CP}.

Let us consider an autonomous dynamical system given by,
\begin{equation}
{\bf\dot x} = {\bf F}({\bf x}),
\label{SYS}
\end{equation}
where ${\bf x}=(x_1,x_2,\dots,x_d)$ is a $d$-dimensional state vector
whose evolution is governed by the function ${\bf F} = (F_1, F_2, \dots,
F_d)$. Given an initial state vector ${\bf x}(0)$ at time $t=0$, the time
evolution ${\bf x}(t)$ is uniquely determined by Eq.~(\ref{SYS}). Now let
us assume that only one component of the state vector is known to us and
we take it to be $x_1(t)$ without loss of generality. The problem that we
address is to obtain the initial state vector ${\bf x}(0)$ from the
knowledge of the scalar signal $x_1(t)$.

Let ${\bf y}(0)$ denote a random initial state vector and ${\bf y}(t)$ its
time evolution obtained from Eq.~(\ref{SYS}). Let ${\bf w}(t)$ denote the
difference
\begin{equation}
{\bf w}(t) = {\bf y}(t) - {\bf x}(t).
\label{DIF}
\end{equation}
We look for the solution of the equation
\begin{equation}
{\bf w}(t) = 0.
\label{wZERO}
\end{equation}
Noting that the initial state vectors ${\bf y}(0)$ and ${\bf x}(0)$
uniquely determine the difference ${\bf w}(t)$, one of the solutions of
Eq.~(\ref{wZERO}) is ${\bf y(0)} - {\bf x}(0)=0$ and this is the solution
that we are searching for.

We now introduce the notation ${\bf w}^n={\bf w}^n({\bf y}^0,{\bf
x}^0)={\bf w}(n\Delta t)$, where $\Delta t$ is a small time interval.
Similarly, ${\bf y}^n = {\bf y}(n\Delta t)$ and ${\bf x}^n = {\bf
x}(n\Delta t)$. With this notation condition~(\ref{wZERO}) can be written
as ${\bf w}^n = 0$.

Our approach to the solution of Eq.~(\ref{wZERO}) is a modified
Newton-Raphson method~\cite{Dev} which includes the time evolution of the
system.

Let us first consider ${\bf w}^1$. We have
\begin{eqnarray}
0 &=& {\bf w}^1({\bf x}^0, {\bf x}^0), \nonumber \\
&=&  {\bf w}^1({\bf y}^0 + \delta {\bf y}^0, {\bf x}^0), \nonumber \\
&=&  {\bf w}^1({\bf y}^0, {\bf x}^0) 
+(\delta {\bf y}^0 \bm{\cdot \nabla}_{{\bf y}^0} )
{\bf w}^1({\bf y}^0,{\bf x}^0) + {\cal O}((\delta {\bf y}^0)^2),
\label{taylorx}
\end{eqnarray}
where $\delta {\bf y}^0 = {\bf x}^0 - {\bf y}^0 = - {\bf w}^0$ and the
last step is a Taylor series expansion in $\delta {\bf y}^0$. For small
$\Delta t$, we can write
\begin{equation}
{\bf w}^1({\bf y}^0, {\bf x}^0)
= {\bf w}^0 + \Delta t \, [{\bf F}({\bf y}^0)-{\bf F}({\bf x}^0)]
+ {\cal O}((\Delta t)^2).
\label{taylort}
\end{equation}
Substituting Eq.~(\ref{taylort}) in Eq.~(\ref{taylorx}) and neglecting
higher order terms, we get,
\begin{equation}
{\bf w}^1({\bf y}^0, {\bf x}^0) = {\bf w}^0 - \Delta t ({\bf w}^0
\bm{\cdot \nabla}_{{\bf y}^0} ) {\bf F}({\bf y}^0).
\label{veceq}
\end{equation}
It is convenient to write the above equation in a matrix form as
\begin{eqnarray}
W^1 &=& (I + \Delta t \, J^0) W^0, \nonumber\\
&=& A^0 W^0,
\label{W1}
\end{eqnarray}
where $W^n$ is the column matrix corresponding to the vector ${\bf w}^n$,
$I$ is the identity matrix, $A^n = I + \Delta t \, J^n$, and the elements
of the Jacobian matrix $J^n$ are given by
\begin{equation}
J^n_{ij} = {\partial F_i({\bf y}^n) \over \partial y_j^n}.
\end{equation}

Next we consider ${\bf w}^2$ or $W^2$. Proceeding as above, we get (see
Eq.~(\ref{W1})),
\begin{eqnarray}
W^2 & = & (I + \Delta t \, J^1) W^1 \nonumber \\
& = & (I + \Delta t \, J^1)(I + \Delta t \, J^0) W^0 \nonumber \\
& = & A^1 A^0 W^0.
\label{W2}
\end{eqnarray}
Similarly, the equation for $W^n$ is
\begin{eqnarray}
W^n & = &  (I+ \Delta t \, J^{n-1})  (I+ \Delta t \, J^{n-2})
\cdots (I+ \Delta t \, J^0) W^0 \nonumber \\
& = & A^{n-1} A^{n-2} \cdots A^0 W^0.
\label{Wn}
\end{eqnarray}

We now concentrate on the first component of the signal whose time series
is assumed to be known. For a $d$-dimensional system we need $d-1$
equations to determine the initial state vector ${\bf x}^0$.
Eqs.~(\ref{W1}), (\ref{W2}) and~(\ref{Wn}) give us the required relations.
\begin{eqnarray}
W^1_1 & = & \sum_{i=1}^d A^0_{1i} W^0_i, \nonumber \\
W^2_1 & = & \sum_{i,j=1}^d A^1_{1i} A^0_{ij} W^0_j, \nonumber \\
\vdots \nonumber \\
W^{d-1}_1 & = & \sum_{i,\ldots,l,m=1}^d A^{d-2}_{1i}  \cdots A^0_{lm} W^0_m,
\label{simeq}
\end{eqnarray}
These are $d-1$ simultaneous equations for $W^0$.

The numerical procedure is as follows. We set the intial state of system
(\ref{SYS}) to a random initial guess vector $\left({\bf
y}^0\right)_{old}$ with $\left(y^0_1\right)_{old}=x^0_1$ and evolve it
using Eq.~(\ref{SYS}). Using this vector ${\bf y}(t)$ we write down $d-1$
simultaneous equations (Eqs.~(\ref{simeq})) which can be solved for $d-1$
unknown components of ${\bf w}^0=-\delta {\bf y}^0$.  Also, $\delta
y^0_1=0$. Thus the initial guess vector can be improved by
\begin{equation} 
\left( {\bf y}^0 \right)_{new} = \left( {\bf y}^0
\right)_{old} + \delta {\bf y}^0. 
\label{ITER} 
\end{equation} 
This sets up an iterative scheme giving us better and better estimates of
the initial vector which converge to ${\bf x}^0$.

We note that as in Newton-Raphson method, the choice of the initial guess
vector can be very important~\cite{PTVF}. In some cases, the iterative
procedure of Eq.~(\ref{ITER}) may not converge or converge to a wrong
root. In such cases, a different choice of initial guess vector can be
useful.

We further note the similarity of our method with the so called method of
variational equations in analytical dynamics~\cite{Whi}. The method of
variational equations can be appied to a known Hamiltonian system to
determine an unknown neighbouring trajectory to an already known one.
There, the method requires a complete particular solution of a known set
of Hamiltonian equations of motion. In contrast, we have used our method
for dissipative chaotic systems. In such systems an analytical solution of
the equations of motion cannot be known. Further our method requires only
one component of a complete trajectory to be sampled. This has important
consequences in the problem of synchronization using a scalar signal.

We now demonstrate our method of estimating the initial state. As our
first example we discuss the R\"ossler system given by~\cite{Ros},
\begin{eqnarray}
\dot x_1 &=& -x_2-x_3,
\nonumber\\
\dot x_2 &=& x_1 + ax_2,
\nonumber\\
\dot x_3 &=& b + x_3(x_1-c).
\label{ROS}
\end{eqnarray}

First, we consider a case when the time series for $x_1$ is given and we
want to estimate $(x^0_2,x^0_3)$. We chose the parameters $(a,b,c)$ such
that the system is in the chaotic regime and the initial state ${\bf x}^0$
is on the chaotic attractor. We start with an arbitrary initial state
${\bf y}^0=(y^0_1,y^0_2,y^0_3)$ with $y^0_1 = x^0_1$. From
Eqs.~(\ref{simeq}) we get a pair of simultaneous equations as,
\begin{eqnarray}
w^1_1 & = y^1_1 - x^1_1 = &  \Delta t \delta y^0_2  + \Delta t \delta y^0_3
\nonumber\\
w^2_1 & = y^2_1 - x^2_1 = &  \left( 2\Delta t+a(\Delta t)^2\right) 
\delta y^0_2 
\nonumber\\
&&+ \left( 2\Delta t+(y^0_1-c)(\Delta t)^2\right) \delta y^0_3
\label{ROSX1}
\end{eqnarray}
which can be solved for $(\delta y^0_2,\delta y^0_3)$. With $\delta
y^0_1=0$ we use these in an iterative manner (Eq.~(\ref{ITER})) to obtain
the correct intial conditions.

Table~1 shows successively corrected $(y^0_2,y^0_3)$ obtained using the
iterative process as discussed. These are the successive estimates for
$(x^0_2,x^0_3)$. Let $e_i=|y^0_i-x^0_i|$ denote the absolute error in the
estimation of $x^0_i$. In Fig.~1(a) we plot a graph corresponding to
Table~1 showing errors $e_2$ and $e_3$ (on logarithmic scale) plotted
against the number of iterations of our method~(Eq.~(\ref{ITER})). From
Table~1 and Fig.~1(a) we see that the successive estimates converge to the
correct values of $(x^0_2,x^0_3)$. Using only two data points in the given
time series $x_1(t)$, we can thus readily estimate the full initial state
${\bf x}^0$. We also note that the rate of convergence is very good. In
about 8 to 10 iterates we obtain the initial values $(x^0_2,x^0_3)$ to
within computer accuracy. If we write the deviations of the succesive
iterates from the correct values in the form 
\begin{equation} 
(e_i)_n = \left| \left(y^0_i\right)_n - x^0_i \right| \sim e^{-\alpha n},
\label{converge} 
\end{equation} 
where $n$ is the number of iterations, then the value of the parameter
$\alpha$ is found to be $2.01$ for $e_2$ and $2.02$ for $e_3$. This is
consistant with the fact that Newton-Raphson method has a quadratic
convergence~\cite{Dev,PTVF}.

We note that the largest Lyapunov exponent for the subsystem
$(y^0_2,y^0_3)$ (conditional or subsystem Lyapunov exponent) is
positive~\cite{PC1}. The success of our method does not depend on whether
this Lyapunov exponent is positive or negative. This is important for
synchronization of chaotic signals as will be discussed afterwards.

We next present cases where time series for the variables $x_2$ and $x_3$
of the R\"ossler system are given. The procedure is similar to the case of
time series for $x_1$ as discussed above.  Fig.~1 (b) shows the errors
$e_1$ and $e_3$, when time series for $x_2$ is given, plotted against the
number of iterations. The parameter $\alpha$ (Eq.~(\ref{converge})) is
$1.97$ for $e_1$ and $1.95$ for $e_3$. This again indicates a quadratic
convergence. Similarly, Fig.~1 (c) shows the quantities $e_1$ and $e_2$
when time series for $x_3$ is given, as a function of the number of
iterations. The parameter $\alpha$ (Eq.~(\ref{converge})) is $1.28$ for
$e_1$ and 1.30 for $e_2$ which shows a convergence slower than quadratic.
We note that the largest subsystem Lyapunov exponent is negative when time
series for $x_2$ is given and is positive when time series for $x_3$ is
given~\cite{PC1}.

As our next example we consider the Chua's circuit which in its
dimensionless form is given by~\cite{CKEI},
\begin{eqnarray}
\dot x_1 &=& \alpha(x_2-x_1-f(x_1)),
\nonumber\\
\dot x_2 &=& x_1-x_2+x_3,
\nonumber\\
\dot x_3 &=& -\beta x_2,
\nonumber\\
f(x_1) &=& bx_1 +{1\over 2}(a-b)\left[|x_1 -1|-|x_1 +1|\right].
\label{CHUA}
\end{eqnarray}
We chose the parameters $a,b,\alpha$ and $\beta$ such that the attractor
is a limit cycle. The initial state ${\bf x}^0$ is chosen in the basin of
attraction of this limit cycle. Fig.~2 shows the errors $e_2$ and $e_3$
when a time series for the variable $x_1$ is given as a function of the
number of iterations of our method. The parameter $\alpha$
(Eq.~(\ref{converge})) in this case takes values $1.29$ for $e_2$ and
$1.25$ for $e_3$ showing a slower than quadratic convergence.

We have also applied our method for the cases when time series for the
variables $x_2$ and $x_3$ are given and also for cases when parameters are
such that the attractor is chaotic. In all the cases we are able to
estimate the full initial state vector.

We have successfully applied our method to estimate the initial state
vector using a given scalar time series for many other dynamical systems
as well. These include Lorenz system~\cite{Lor} in its periodic, chaotic
or intermittent regimes, the disk dynamo system modelling a periodic
reversal of earth's magnetic field~\cite{Rik,FW}, a 3-d plasma system
formed by a three wave resonant coupling equations~\cite{WFO} and a four
dimensional phase converter circuit~\cite{YK}.

Now we will discuss an important application of our estimation method in
the problem of synchronization of two identical chaotic systems coupled
unidirectionally by a scalar signal. Let us suppose that Eq.~(\ref{SYS})
describes a chaotic system and let us consider a replica of it given by,
${\bf\dot y} = {\bf F}({\bf y})$.  Without losing generality let's further
assume that a scalar output signal $x_1(t)$ is given. The aim is to
synchronize vector ${\bf y}(t)$ with ${\bf x}(t)$ using this scalar
signal. Our method of estimating initial conditions makes this procedure
trivial. Using the scalar signal we estimate the initial vector ${\bf
x}^0$ and set ${\bf y}^0 = {\bf x}^0$. This clearly leads to an
instantaneous synchronization of the two trajectories. As we have
demonstrated in the case of R\"ossler system, our method works even when
the largest conditional Lyapunov exponent is positive which is the case
when other methods of synchronization are known to fail~\cite{PC1,CP}.

To summarise, we have introduced a novel yet simple method to estimate
initial conditions of a multivariable dynamical system from a given scalar
signal. Our method is based on a multidimensional Newton-Raphson method
where we include the time evolution of the system. The method gives a
reasonably fast convergence to the correct initial state. The required
length of the time series is very small. The method works even when the
largest conditional Lyapunov exponent is positive. An important
consequence of the method is that the problem of synchronization of
identical chaotic systems using scalar signal becomes trivial since
evolution of two such systems can then be started from identical initial
states.

\end{multicols}

\begin{table} 
\caption{The table presents successive estimates of the unknown
components of the initial state vector, i.e. $(y^0_2,y^0_3)$, in a
R\"ossler system for which a time series $x_1(t)$ is given. $n$ is the
number of iterations of our method~(Eq.~(\ref{ITER})).  A clear
convergence is seen towards the actual values $x^0_2,x^0_3$, written at
the bottom of the table. The parameters are $(a,b,c)=(0.2,0.2,9.0)$. The
time step $\Delta t$ is 0.01 and we use fourth order Runge-Kutta method
for time evolution of Eq.~(\ref{ROS}).  The corresponding plots are shown
in Fig.~1(a).}
\begin{tabular}{ccc}
$n$ & $y^0_2$ & $y^0_3$\\
\hline
         0   &    $-9.168384973984731$  &      7.4923642142341230\\  
         1   &    $-5.166404733407783$  &     0.4244936662730687\\
         2   &    $-5.178027081551812$  &     0.4761293644577211\\   
         3   &    $-5.178176377100173$  &     0.4759664825328775\\
         4   &    $-5.178173198394843$  &      0.4759644699505680\\
         5   &    $-5.178173232249095$  &     0.4759645133108157\\
         6   &    $-5.178173232007102$  &     0.4759645128302578\\
         7   &    $-5.178173232003735$  &     0.4759645128297332\\
         8   &    $-5.178173232003735$  &     0.4759645128297332\\
\hline
     & $x^0_2=-5.178173232007801$ &  $x^0_3=0.4759645128337372$\\
\end{tabular}
\end{table}

\begin{figure}
\epsffile{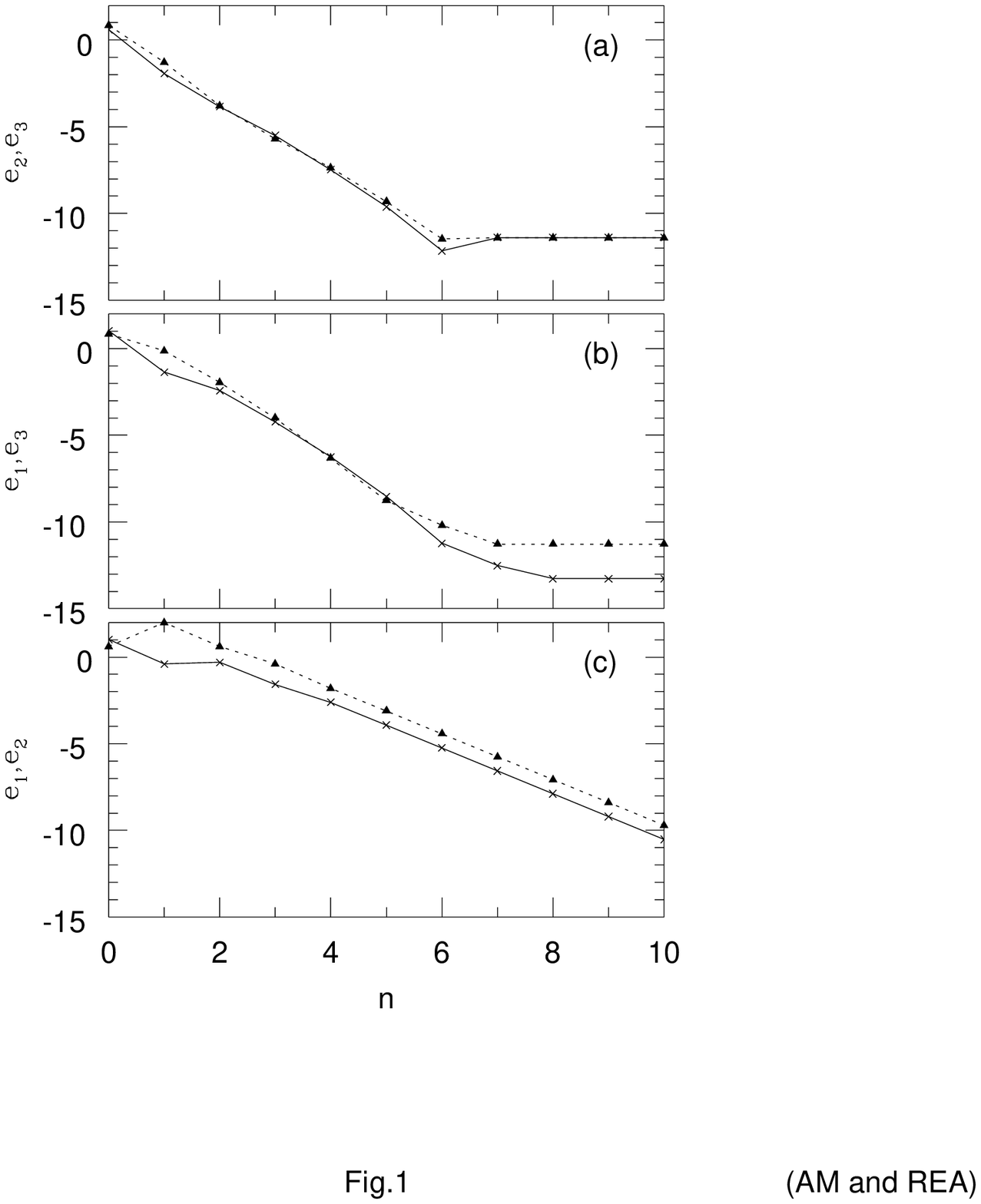}
\caption{ Plot (a) shows the errors $e_2$ and $e_3$ on a logarithmic
scale, as a function of $n$, the number of iterations of our method for
R\"ossler system when time series for variable $x_1$ is given. The crosses
show the values for $e_2$ and solid triangles those for $e_3$. Errors are
seen to approach zero as $n$ increases. The values are taken from Table~1.
Similarly plot (b) shows the errors $e_1$ (crosses), and $e_3$ (solid
triangles) as a function of $n$ when the time series for $x_2$ is given.
Plot (c) shows the errors $e_|$ (crosses), and $e_2$ (solid triangles) as
a function of $n$ when the time series for $x_3$ is given. For all these
cases, the parameters are such that the attractor is chaotic and the
initial state is on the attractor.} \end{figure}

\begin{figure}
\epsffile{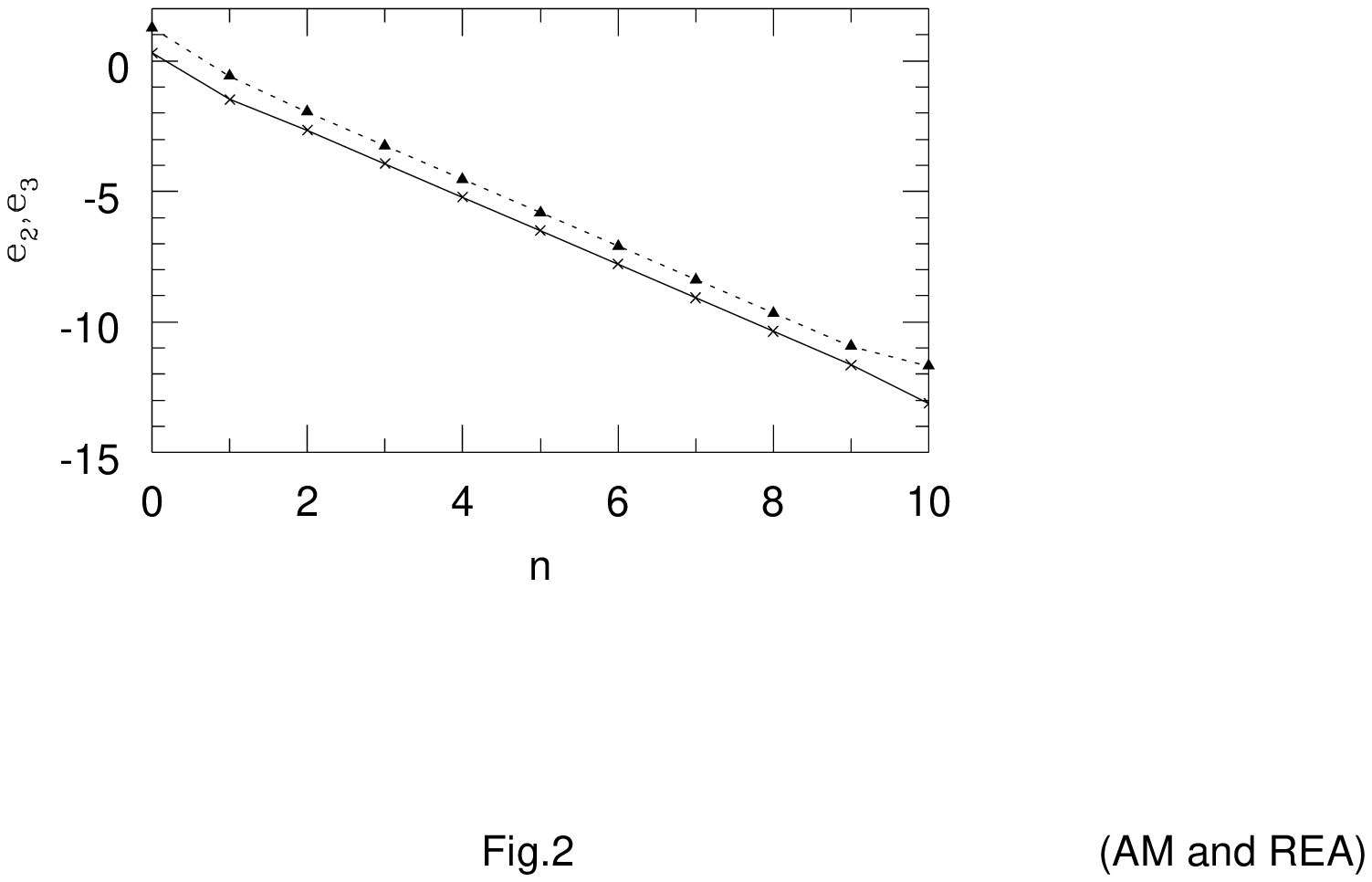}
\caption{ The figure shows errors $e_2$ and $e_3$ on a
logarithmic scale, as a function of $n$, the number of iterations of our
method for Chua's circuit when time series for the variable $x_1$ is
given. The crosses show the values for $e_2$ and solid triangles those for
$e_3$. Errors are seen to approach zero as $n$ increases. The parameters
are$(4,3,2,0.1758)$ and are such that the attractor is a limit cycle and
the initial state is in the basin of the attractor. } \end{figure}


\begin{references}

\bibitem{Dev} R.L. Devaney, {\it An Introduction to Chaotic Dynamical
Systems}, (The Benjamin / Cummings Pub. Co. Inc., Menlo Park - California,
1986).

\bibitem{PTVF} W.H. Press, S.A. Teukolsky, W.T. Vetherling and B.P.
Flannery, {\it Numerical Recipes in C}, 2nd ed. (Cambridge University, New
York, 1992), p. 379.

\bibitem{ABST} H.D.I. Abarbanel, R. Brown, J.J. Sidorowich, L.Sh.
Tsimring, Rev. of Mod. Phys. {\bf 65}, 1331 (1993).

\bibitem{PC1} L.M. Pecora and T.L. Carroll, Phys. Rev. Lett. {\bf 64}, 821
(1990).

\bibitem{CO} K.M. Cuomo and A.V. Oppenheim, Phys. Rev. Lett. {\bf 71}, 65
(1993).

\bibitem{PKSP} U. Parlitz, L. Kocarev, T. Stojanovski, H. Preckel, Phys.
Rev. E {\bf 53}, 4351 (1996) and references therein.

\bibitem{JA} J. K. John and R.E. Amritkar, Int. J. Bifurcation Chaos {\bf
4}, 1687 (1994).

\bibitem{Par} U. Parlitz, Phys. Rev. Lett. {\bf 76}, 1232 (1996).

\bibitem{MA} A. Maybhate and R.E. Amritkar, Phys. Rev. E {\bf 59}, 284
(1999).

\bibitem{CP} T.L. Carroll and L.M. Pecora in {\it Nonlinear dynamics in
circuits}, ed. L.M. Peccora and T.L. Carroll (World Scientific Pub. Co.
Pte. Ltd., Singapore, 1995) p. 215.

\bibitem{Whi} E. T. Whittaker {\it A treatise on the analytical dynamics
of particles and rigid bodies}, $4^{th}$ Ed. (Dover publications, NY,
1944) p. 268.

\bibitem{Ros} O.E. R\"ossler, Phys. Lett. A {\bf 57}, 397 (1976).

\bibitem{CKEI}L.O. Chua, L. Kocarev, K. Eckart and M. Itoh, Int. J.
Bifurc. and Chaos {\bf 2}, 705 (1992); also see Ref.~[6] p.238.

\bibitem{Lor} E.N. Lorenz, J. Atmos. Sci. {\bf 20}, 130 (1963).

\bibitem{Rik} T. Rikitake, Proc. Cambridge Philos. Soc. {\bf 54}, 89
(1958).

\bibitem{FW} C. Flynn and N. Wilson, Am. J. Phys. {\bf 66}(8), 730 (1998).

\bibitem{WFO} J.M. Wersinger, J.M. Finn, E. Ott, Phys. Rev. Lett. {\bf
44}, 453 (1980).

\bibitem{YK} T. Yoshinaga and H. Kawakami in {\it Nonlinear dynamics in
circuits}, ed. L.M. Peccora and T.L. Carroll (World Scientific Pub. Co.
Pte. Ltd., Singapore, 1995) p. 111.

\end{references}
\end{document}